\documentclass[aps,pra,reprint,showpacs,notitlepage,superscriptaddress,twocolumn]{revtex4-2}
\usepackage[utf8]{inputenc}
\usepackage{float}
\usepackage{amssymb}
\usepackage{mathrsfs}

\usepackage{gensymb}
\usepackage{braket}

\usepackage{amsfonts}
\usepackage{graphicx}
\usepackage{amsmath}
\usepackage{dcolumn}
\usepackage{color}
\usepackage{subfigure}
\usepackage{epsfig}
\usepackage{soul}

\usepackage[usenames,dvipsnames]{xcolor}

\usepackage[colorlinks,linkcolor=blue,citecolor=blue,hyperindex,bookmarks=false,pdfstartview=FitH]{hyperref}

\usepackage{footmisc}
\usepackage{lipsum}

\usepackage{bm}
\usepackage{dsfont}
\usepackage{verbatim}
\usepackage{multirow}
\usepackage{mathrsfs}
\usepackage{leftidx}
\usepackage{xspace}
\usepackage{braket}
\usepackage{bbm}
\usepackage{cancel}
\usepackage[normalem]{ulem}

\usepackage{mleftright} 
\usepackage{siunitx}

\providecommand{\U}[1]{\protect\rule{.1in}{.1in}}

\newcommand{\figpanel}[2]{\hyperref[#1]{\ref*{#1}(#2)}}

\definecolor{alanred}{RGB}{102, 0, 255}

\begin{document}

\title{Tunable dual-band atomic mirror based on subwavelength atomic arrays under electromagnetically induced transparency}

\author{Shiwen Sun}
\affiliation{School of Physics and Center for Quantum Sciences, Northeast Normal University, Changchun 130024, China}
\affiliation{These authors contributed equally to this work.}

\author{Yi-Xin Wang}
\affiliation{School of Physics and Center for Quantum Sciences, Northeast Normal University, Changchun 130024, China}
\affiliation{These authors contributed equally to this work.}

\author{Xiao Liu}
\affiliation{College of Physical Science and Technology, Heilongjiang University, Harbin 150080, China}

\author{Yan Zhang}
\email{zhangy345@nenu.edu.cn}
\affiliation{School of Physics and Center for Quantum Sciences, Northeast Normal University, Changchun 130024, China}

\date{\today }

\begin{abstract}
Subwavelength atomic arrays offer a powerful platform for engineering cooperative light-matter interactions and enabling quantum metasurfaces.
We demonstrate that a two-dimensional array of three-level atoms operating under electromagnetically induced transparency can function as a tunable dual-band atomic mirror, where two independently controllable reflection bands emerge from the collective optical responses mediated by dipole-dipole interactions.
These resonances yield dual reflection bands with asymmetric linewidths, whose spectral positions and bandwidths can be tuned through the control-field parameters, dipole orientation, incident geometry, and lattice constant.
We further identify the conditions under which additional diffraction orders emerge, which delineate the operational and tunable range of the atomic mirror via its collective-mode structure.
This scheme provides a fully tunable dual-band atomic mirror operating across broad frequency and angular ranges, offering a practical and experimentally accessible pathway toward reconfigurable photonic elements in atomic-array platforms at low energy levels.

\end{abstract}

\maketitle

\section{Introduction}

Photonics drives our modern information society, whereas electronic technologies face bandwidth and energy efficiency challenges due to exponential traffic growth~\cite{Sinence2010}.
Photons have the merits of high data rates, low power consumption, broad bandwidth, and decoherence resistance, making them ideal information carriers~\cite{PRL2024Photon-encoding, PhotonRes2022}.
Thus, shaping and controlling photon flows is crucial in optical networks, driving research in photonic crystals~\cite{PC1987PRL,  ZhangYOE2011}, chiral phase modulation~\cite{WangYX2023PRA}, unidirectional reflection~\cite{Liuxiao2024NJP, ZhangYOE2021, NatPho2015KK, WuJH2014PRL}, nonreciprocal scattering~\cite{Liuxiao2025OE, Nonreciprocal2018, NonreciprocalPRL2018}, and unidirectional amplification~\cite{UAPRL2018, UAPRL2019, UACommPhys2022}.
As a fundamental quantum interference effect, electromagnetically induced transparency (EIT) provides a versatile method for all-optical control, offering tunable transparency and slow-light behavior with minimal absorption~\cite{EIT2005RMP}.
These features enable a wide range of applications, including slow/fast light~\cite{PhysRevA.83.013827, PhysRevA.96.053823}, photonic band-gap engineering~\cite{PhysRevA.91.013826,   PhysRevA.94.013836, Yutong2025, Muhua2025}, optical switching~\cite{albert2011cavity}, and light storage~\cite{PhysRevLett.111.033601}.
Additionally, the nonlinear response and integration of materials like graphene~\cite{grapheneNatPho2010, grapheneNatNan2009, grapheneNature2011}, active media~\cite{Active1986, Active2003}, and semiconductor photonic crystal cavities~\cite{Semiconductor2010} further enhance photon flow control.

Metasurfaces, with their ability to precisely engineer subwavelength patterns from dielectric or metallic thin films, have emerged as a novel platform for manipulating photons and designing optical networks~\cite{QMScience, QMNat, QMSci2017, Jingwen2026}.
Subwavelength atomic arrays, recently referred to as atomic metasurfaces, represent a new paradigm for quantum metamaterials~\cite{NatPhys2020Engtaglement}.
With the fully coherent dipole-dipole interactions between atoms, the collective lattice modes significantly enhance the atomic optical cross section, thereby facilitating an efficient light-atom interface~\cite{Top2, Scs3, OE2022ring}.
These metasurfaces also exhibit a dipole-blocking effect, which, along with their exceptional non-classical light scattering properties, distinguishes them from Rydberg atom-based systems~\cite{PhBlo1, PhBlo2}.
In radiative dynamics, this system allows photons to propagate within the material with an extremely long lifetime, which is crucial for efficient photon storage~\cite{NonMultiPhotonPRA2024, LightstorePRL2016, PRX2017} and transfer~\cite{2025NJPtransferstates, PRR2024transport}.
This phenomenon has sparked interest in exploring its potential applications in one-dimensional atomic-waveguide quantum electrodynamics (QED), aiming to mitigate the wave losses typically encountered in traditional waveguides~\cite{AQED1, AQED2, AQED3}.
This concept has been further extended to the two-dimensional (2D) configuration~\cite{2DAQED}.
Notably, by designing the topological distribution and optimizing the internal degrees of freedom of atoms, metasurface materials can be engineered to support robust long-lived topological modes~\cite{Top1, Top2, Top3}, including high-order topological photon states~\cite{wang2025}.
As a burgeoning class of quantum metamaterials, atomic metasurfaces have been the subject of various studies investigating their application prospects, such as entangled states in quantum networks~\cite{PRL2024MLEntanglement, NatPhys2020Engtaglement}, cooperative sensing~\cite{PRL2024ChiralSening, PRA2025sensing}, chirality-induced spin-orbit coupling~\cite{PRA2024SOcoupling}, and collectively enhanced ground-state cooling~\cite{GSCPRA2025}.
The properties of these materials continue to drive research and development in quantum optics and photonics.

Recently, subwavelength atomic arrays have garnered significant attention in the field of optical engineering, offering promising applications in wavefront engineering~\cite{Wavefront2021} and optical magnetic mirrors~\cite{OMR1, Pem2}.
A recent experimental breakthrough demonstrated that a monolayer atomic metasurface can function as an atomic mirror, enabling effective manipulation of light propagation~\cite{NatureAtomMirror2020}. Additionally, by leveraging Rydberg interactions in conjunction with subwavelength atomic arrays, optically-controllable switches and coherent photon-photon gates can be realized via the Rydberg blockade effect~\cite{PRL2021, Quantum2022}, experimentally validated in recent studies~\cite{NatPhys2023}.
However, this approach is constrained by the interplay between the Rydberg blockade radius and the beam waist, while a single cooperative resonance peak limits the operational bandwidth in optical network applications.
Consequently, broadening the spectral response range and developing advanced control strategies remain critical challenges in atomic mirror technologies.

To overcome the limitations discussed above, we propose an alternative approach that realizes a dual-band atomic mirror based on a subwavelength atomic array operating in the EIT regime without Rydberg states.
The present model enables flexible control over the mirror characteristics, such as the bandwidth and operating frequency range, by tuning the system’s field parameters within experimentally feasible regimes.
Moreover, the polarization-dependent response of the array provides a versatile platform for implementing polarization-selective functionalities, including beam splitting and filtering.
The structure of this paper is organized as follows: In Sec.~\ref{II}, we present the system model and scattering theory, explaining the calculations.
In Sec.~\ref{III}, we investigate the directional collective modes and the cooperative optical response of the system to the incident probe field, where the response function is effectively described by decaying-dressed states.
In Sec.~\ref{IV}, we analyze the reflection and transmission spectra of the system to demonstrate the behavior of the atomic mirror, supported by an examination of the first-order diffraction.
Finally, in Sec.~\ref{V}, we summarize our findings and conclusions.

\begin{figure}[t]
\centering
\includegraphics[width=0.5\textwidth]{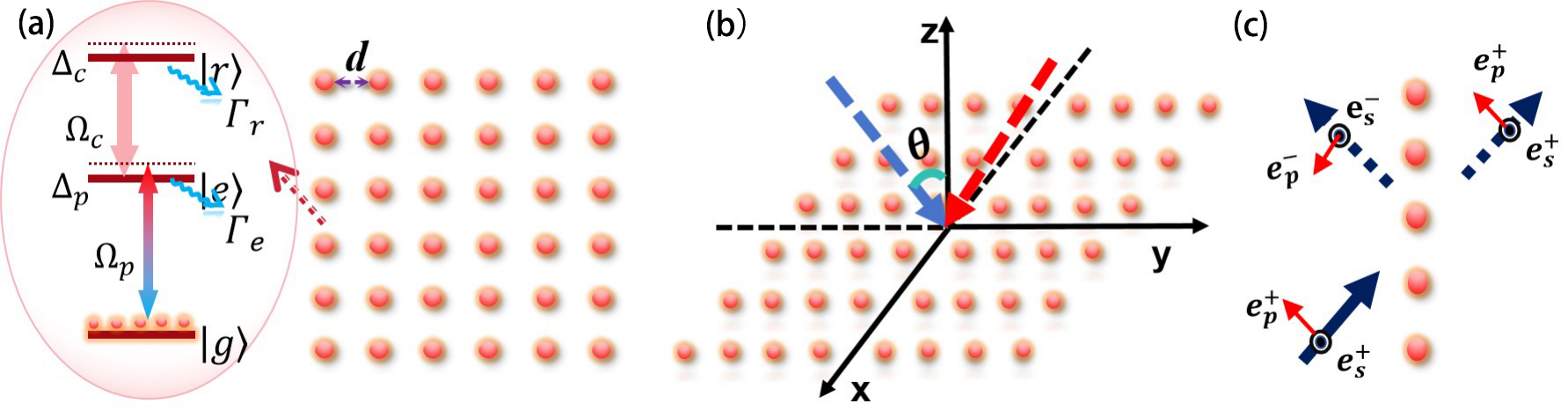}
\caption{(a) A square atomic array positioned in the $x$-$y$ plane at $z=0$ in free space, where each atom is driven into a three-level system.
(b) Schematic of two types of oblique incidence for the weak probe field with an incident angle $\theta$, where the red (blue) arrow represents incident in the $x$-$z$ ($y$-$z$) plane with a wavevector component $k'_{y}=0$ ($k'_{x}=0$).
(c) Scattering of the probe field at a general incidence angle, illustrating the polarization states of the incident and scattered fields within the orthogonal polarization basis $\{\mathbf{e}_{p}^{\pm}, \mathbf{e}_{s}^{\pm}\}$, where the superscript $+$ ($-$) corresponds the transmitted (reflected) field.
}
\label{fig:1}
\end{figure}

\section{Model and equations}\label{II}

As shown in Fig.~\figpanel{fig:1}{a}, we consider a 2D square atomic array with lattice constant $d$ in the $x$-$y$ plane, where each atom is driven into a three-level ladder-type configuration with a ground state $|g\rangle$ and two excited states $|e\rangle$ and $|r\rangle$.
These states correspond to the energy levels of the $^{87}\mathrm{Rb}$ atom, specifically $|g\rangle=\left| 5S_{1/2}, F=2\right\rangle$, $|e\rangle=\left| 5P_{1/2}, F=3\right\rangle$, and $|r\rangle=\left| 7S_{1/2}, F=3 \right\rangle$.
A strong control field with amplitude $\boldsymbol{E}_{c}$ and frequency $\nu_c$ drives the transition $|r\rangle \leftrightarrow |e\rangle$, characterized by Rabi frequency $\Omega_c=\frac{1}{ 2\hbar} \boldsymbol{\mu}_{re}\cdot \boldsymbol{E}_{c} $ and detuning $\Delta_c=\nu_c-\omega_{re}$. Here, $\boldsymbol{\mu}_{ij}$ and $\omega_{ij}$ ($i,j=g,e,r$) denote the electric-dipole moment and the resonant frequency of the corresponding atomic transition, respectively.
Meanwhile, a weak probe field with amplitude $\boldsymbol{E}_{p}$ and frequency $\nu_p$ drives $|e\rangle \leftrightarrow |g\rangle$, with Rabi frequency $\Omega_p= \frac{1}{ 2\hbar}  \boldsymbol{\mu}_{eg}\cdot \boldsymbol{E}_{p}$, detuning $\Delta_p=\nu_p-\omega_{eg}$, and wavevector $k'$.

Under the electric-dipole and rotating-wave approximations, the real-space non-Hermitian effective Hamiltonian is given by
\begin{align}
H_{\rm eff}/\hbar&=\sum_{j=1}^{N} \left[ \left(\omega_{e}-i\frac{\Gamma_{e}}{2}\right) \hat{\sigma}_{ee}^{j}+ \left(\omega_{r} -i\frac{\Gamma_{r}}{2}\right) \hat{\sigma}_{rr}^{j} \right]  \nonumber\\
&- \sum_{j=1}^{N} \left(\Omega_{p}\hat{\sigma}_{eg}^{j}e^{-i\nu_{p} t}+\Omega_{c}\hat{\sigma}_{re}^{j}e^{-i\nu_{c}t}+H.c\right) \nonumber\\
&-\frac{3\pi\Gamma_e c}{\omega_{eg}} \sum_{j\neq i}^N \boldsymbol{\wp_{i}}^{*}\cdot \boldsymbol{G}(\boldsymbol{r}_i,\boldsymbol{r}_j) \cdot\boldsymbol{\wp_{j}} \hat{\sigma}_{eg}^{i}\hat{\sigma}_{ge}^{j} ,
\label{H_eff}
\end{align}
where $N$ is the number of atoms in the array, $c$ is the speed of light, $\boldsymbol{\wp}_j=\boldsymbol{\wp}=\boldsymbol{\mu}_{eg}/|\boldsymbol{\mu}_{eg}|$ is the unit vector of atomic polarization.
Here, we assume that all array atoms have identical polarization, $\boldsymbol{r}_i$ is the position of the $i$th atom, and $\Gamma_{e(r)} = |\boldsymbol{\mu}_{eg(re)}|^2 \omega_{eg(re)}^3 / (3\pi\epsilon_0\hbar c^3)$ is the radiation linewidth of an individual atom in free space for the level $|e\rangle$ ($|r\rangle$).
Unlike traditional EIT schemes~\cite{ZhangYOE2011, WangYX2023PRA, Liuxiao2024NJP, ZhangYOE2021}, the final term of Eq.~(\ref{H_eff}) represents the dipolar spin-spin interaction that can be captured by the dyadic Green's function $\boldsymbol{G}$ in free space~\cite{1998Green}.
This is derived by adiabatically eliminating the reservoir degree of freedom in the Born-Markov approximation and cannot be neglected at subwavelength scales.

In the thermodynamic limit ($N\rightarrow\infty$), the Hamiltonian $H_{\rm eff}$, within the single-excitation subspace, can be transformed into momentum space using the Fourier transform.
The Bloch operators $S_{\mathbf{k}}^{\dagger}$ and $M_{\mathbf{k}}^{\dagger}$ are defined as
\begin{equation}
    \begin{split}
    S_{\mathbf{k}}^{\dagger}&=\frac{1}{\sqrt{N}}\sum_{j}\sigma_{eg}^{j}e^{i\mathbf{k}\cdot \mathbf{r}_{j}}, \\
    M_{\mathbf{k}}^{\dagger}&=\frac{1}{\sqrt{N}}\sum_{j}\sigma_{re}^{j}e^{i\mathbf{k}\cdot \mathbf{r}_{j}}.
    \end{split}
\end{equation}
The Hamiltonian in the interaction picture can be expressed in momentum space as
\begin{align}
H_{\rm eff}/\hbar=&\sum_{\mathbf{k}} \mathcal{H}_{\mathbf{k}}/\hbar \nonumber\\
=&\sum_{\mathbf{k}} \left[ -\left( \Omega_{p,\mathbf{k}}S_{\mathbf{k}}^{\dagger}+\Omega_{c}M_{\mathbf{k}}^{\dagger}+c.c \right)
  \right.\nonumber\\
  &\left.+\left(\xi-\Delta_{p}\right)M_{\mathbf{k}}^{\dagger}M_{\mathbf{k}}+\left(\eta-\Delta_{p}\right)S_{\mathbf{k}}^{\dagger}S_{\mathbf{k}} \right],
  \label{H_k}
\end{align}
where $\Omega_{p,\mathbf{k}}=\frac{1}{ 2\hbar} \boldsymbol{\mu}_{eg} \cdot \boldsymbol{E}_{p,\mathbf{k}}$ with $\boldsymbol{E} _{p,\mathbf{k}} = \frac{1}{\sqrt{N}}   \sum_{j} \boldsymbol{E}_{p} e^{-i\mathbf{k}\cdot\mathbf{r}_{j}}$,
$\xi=-\Delta_c-i\Gamma_r/2$, and $\eta=\Delta_{\mathbf{k}} - i\Gamma_{\mathbf{k}}/2$ with cooperative shift $\Delta_{\mathbf{k}}$ and cooperative decay rate $\Gamma_{\mathbf{k}}$.
Here, $\eta = -3\pi\Gamma_e c \boldsymbol{\wp}^{*}\cdot G_{\mathbf{k}}\cdot\boldsymbol{\wp}/\omega_{eg}  -i\Gamma_e/2$~\cite{Top1, Top2},
where $G_{\mathbf{k}}=\sum_{i\neq j} \boldsymbol{G}(\boldsymbol{r}_i,\boldsymbol{r}_j)  e^{-i\mathbf{k}\cdot(\mathbf{r}_{i}-\mathbf{r}_{j})}$ is the discrete Fourier transform of the free-space tensor.

The master equation of the density operator $\rho_{\mathbf{k}}$ in momentum space is given by $\dot{\rho_{\mathbf{k}}} = -\frac{i}{\hbar} [\mathcal{H}_{\mathbf{k}},\rho_{\mathbf{k}}]+ \Gamma_{\mathbf{k}} \hat{S}_{\mathbf{k}}\rho_{\mathbf{k}}\hat{S}_{\mathbf{k}}^{\dagger} +\Gamma_r\hat{M}_{\mathbf{k}}\rho_{\mathbf{k}}\hat{M}_{\mathbf{k}}^{\dagger}$.
Then we obtain the motion equations
\begin{equation}
\begin{aligned}
\frac{\partial}{\partial t} \rho_{eg,\mathbf{k}}
&= -i\left[ \Omega_{p,\mathbf{k}}\left(\rho_{ee,\mathbf{k}} - \rho_{gg,\mathbf{k}}\right) + \left( \eta-\Delta_p \right)\rho_{eg,\mathbf{k}}-\Omega_{c}\rho_{rg,\mathbf{k}}\right]\\
\frac{\partial}{\partial t} \rho_{rg,\mathbf{k}} &= -i\left[ -\Omega_{c}\rho_{eg,\mathbf{k}}  + \Omega_{p,\mathbf{k}}\rho_{re,\mathbf{k}}+\left(\xi-\Delta_{p}\right)\rho_{rg,\mathbf{k}}\right]\\
\frac{\partial}{\partial t} \rho_{re,\mathbf{k}}&=-i\left[\Omega_{c}\left(\rho_{rr,\mathbf{k}} - \rho_{ee,\mathbf{k}}\right)+\left(\xi-\eta\right)\rho_{re,\mathbf{k}}+\Omega_{p,\mathbf{k}}\rho_{rg,\mathbf{k}}\right]
\\
\frac{\partial}{\partial t} \rho_{gg,\mathbf{k}}&=-i\left[-\Omega_{p,\mathbf{k}}^{*}\rho_{eg,\mathbf{k}}+\Omega_{p,\mathbf{k}}\rho_{ge,\mathbf{k}}\right]+\Gamma_{\mathbf{k}}\rho_{ee,\mathbf{k}}
\\\frac{\partial}{\partial t} \rho_{ee,\mathbf{k}}&=-i\left[-\Omega_{p,\mathbf{k}}\rho_{ge,\mathbf{k}}+\Omega_{p,\mathbf{k}}^{*}\rho_{eg,\mathbf{k}}+\Omega_{c}\rho_{er,\mathbf{k}}-\Omega_{c}^{*}\rho_{re,\mathbf{k}}\right]\\
&~~~~+\Gamma_{r}\rho_{rr,\mathbf{k}}-\Gamma_{\mathbf{k}}\rho_{ee,\mathbf{k}}
\end{aligned}
\end{equation}
constrained by $\rho_{ij,\mathbf{k}}=\rho_{ji,\mathbf{k}}^*$ and $\sum\rho_{jj}=1$.

In the weak-probe limit, the complex probe susceptibility of a three-level system is written as
\begin{align}
\chi_{\rm eff}=\frac{\left | \boldsymbol{\mu}_{eg} \right| ^2 }{\hbar}\frac{\rho_{eg,\mathbf{k}}}{\Omega_{p,\mathbf{k}}}=\frac{\left | \boldsymbol{\mu}_{eg}  \right | ^{2}/\hbar}{\eta-\Delta_{p}-\frac{\left | \Omega _{c}  \right |^{2}  }{\xi-\Delta_p  }} .
\label{Eq:3}
\end{align}
This linear response has two poles
\begin{align}
    \Delta_{\pm}=(\xi+\eta\pm\sqrt{(\xi -\eta)^{2}+4\left|\Omega_{c}\right|^{2}})/2
\end{align}
These poles exhibit distinct resonance responses regarding the effective states, i.e., decaying-dressed states~\cite{anisimov2008decaying}, which originate from the interaction between the usual dressed states (i.e., the eigenvalues of the 'atom+field' Hamiltonian) and the environment.
The real and imaginary parts of these poles determine the frequencies and dephasing rates of the two effective states, respectively.

According to the method described in Ref.~\cite{PRL2017scattering}, the scattering coefficients of the probe field $\boldsymbol{E}_{p}$ can be accurately captured.
The generalized input-output equation of the probe field reads
\begin{align}
\mathbf{E}(\mathbf{r}) = \left[ e^{ik'_z z} + S_{\pm}(\mathbf{k'}_{\parallel}) \, e^{ik'_z |z|} \right] e^{i\mathbf{k'}_{\parallel} \cdot \mathbf{r}_{\parallel}} \mathbf{E}_{p},
\label{eq:scattering_field}
\end{align}
with the scattering matrix
\begin{align}
S_{\pm}(\mathbf{k'}_{\parallel})
= \frac{k'^2}{\varepsilon_0}
e^{-i\mathbf{k'}_{\parallel}\cdot\mathbf{r}_{\parallel}}
e^{-ik'_z|z|}G_{\mathbf{k'}_{\parallel}}
\chi_{\rm eff}.
\label{eq:scattering_matrix}
\end{align}
Here, $k'_{x,y,z}$ represent the components of the incident wave vector, $k'_{\parallel}=\sqrt{{k'}_x^2+{k'}_y^2}$ represents the magnitude of the wave vector in the plane of the array, and the superscript $\pm$ denotes the direction of propagation relative to the $z$-axis, as shown in Fig.~\figpanel{fig:1}{c}.
Within the $s$-$p$ polarization basis, the incident field can be expressed in terms of the corresponding unit vectors $\mathbf{e}_{p,s}$.
These vectors are defined as
\begin{align}
   \mathbf{e}_{p}^{\pm } &=\pm \frac{k'_{z}}{k' k'_{\parallel} } \left ( k'_{x}, k'_{y},\mp  {k'}_{\parallel} ^{2}/ k'_{z}\right), \\
  &~~~\mathbf{e}_{s}^{\pm } =\frac{1}{k'_{\parallel } } \left ( k'_{y} ,-k'_{x},0 \right).
\end{align}
The optical response of the system is characterized by the scattering matrix, which is a $2\times2$ matrix expressed in two orthogonal polarization basis.
The matrix elements are given by
\begin{align}
    S_{\mu \nu }^{\pm } =\frac{i\pi k'}{d^{2}\varepsilon _{0}\lambda k'_{z}}e_{\mu }^{\pm } \chi_{\rm eff} e_{\nu }^{+}
    \label{S_pm}
\end{align}
where $\mu,\nu=p,s$ represents the basis polarization state of the probe field.
Then the transmission and reflection of the corresponding polarized light are denoted as $T_{\mu \nu } =\left | \delta _{\mu \nu }  +S_{\mu \nu }^{+} \right |^{2}$ with the Kronecker delta $\delta_{\mu\nu}$ and $ R_{\mu \nu } =\left |S_{\mu \nu }^{-} \right |^{2}$, respectively.

\section{Collective responses of the atomic array in EIT}\label{III}

\begin{figure}[t]
\centering
\includegraphics[width=0.48\textwidth]{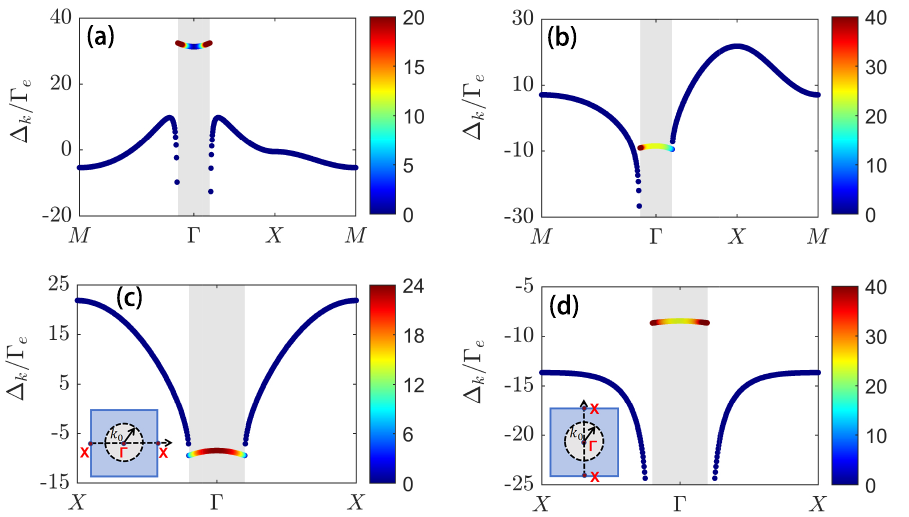}
\caption{Band structure of the atomic array for (a) out-of-plane polarization $\hat{\wp}= \hat{\wp_{z}}$, and (b) in-plane polarization $\hat{\wp}= \hat{\wp_{x}}$.
Band structure with in-plane polarization $\hat{\wp}= \hat{\wp_{x}}$ along the Bloch vector (c) $k_y=0$ path, and (d) $k_x=0$ path.
The insets indicate the specific path in the first Brillouin zone.
$\Gamma$, $X$ and $M$ are
the symmetry points.
The collective decay rate $\Gamma_{k}$ is color-coded.
The shade indicates the light cone $|\mathbf{k}_{\parallel}|=\omega_{eg}/c$.
The parameters are $\lambda=2\pi c/\omega_{eg}=790$ nm, $\Gamma_e=2\pi\times6$~MHz,  $\Gamma_r = 2\pi \times 1.8~\text{MHz}$ and $d=0.1\lambda$.}
\label{fig:2}
\end{figure}

As demonstrated by Eq.~\ref{Eq:3}, the collective atomic mode of the system induces a frequency shift in the excited state $ |e\rangle$ and modifies its radiative bandwidth, thereby significantly influencing the optical response.
Figures~\figpanel{fig:2}{a}--\figpanel{fig:2}{b} show the cooperative shift $\Delta_{k}$ and collective decay rate $\Gamma_{k}$ for out-of-plane atomic polarization $\hat{\wp}=\hat{\wp}_{z}$ and in-plane atomic polarization $\hat{\wp}=\hat{\wp}_{x}$, respectively, along a high-symmetry path of the first Brillouin zone.
In both polarization cases, the collective mode exhibits strong coupling (sufficiently large $\Gamma_{k}$) to free-space modes within the light cone $(kc<\omega_{eg})$, manifesting superradiant characteristics due to constructive interference between array atoms.
Note that, for out-of-plane case, the system displays perfect subradiant characteristics at the $\Gamma$ point, i.e., $\Gamma_{\mathbf{k}}=0$ (see Fig.~\figpanel{fig:2}{a}).
This arises because radiation along the orientation of the atomic dipole polarization is forbidden, which also leads to a strong suppression of the collective decay rate for other modes near this point~\cite{PRX2017}.
In contrast, atomic modes outside the light cone $(kc>\omega_{eg})$ are entirely decoupled from free-space photons due to quasimomentum mismatch, resulting in subradiant states with negligible $\Gamma_{\mathbf{k}}$.

For the in-plane polarization, as shown in Fig.~\figpanel{fig:2}{c}--\figpanel{fig:2}{d}, we vary the path of the Bloch vector within the Brillouin zone and observe that the energy band structure of the array undergoes significant variations depending on the chosen path.
In contrast, the collective response of the atomic array for out-of-plane polarization remains largely insensitive to the path of the Bloch vector (see Fig.~\figpanel{fig:2}{a}).
The primary reason for this distinction lies in the fact that these systems belong to different space point groups.
Specifically, the in-plane polarization atomic array is protected by the space point group $C_{2v}$, while the out-of-plane polarization array is protected by $C_{4v}$.

The response of the atomic array is also influenced by the incident orientation of the probe field $\mathbf{E}_p$.
This dependence arises because collective modes within the light cone are excited when the projection components $k'_x$ and $k'_y$ of the incident wave vector $k' \approx k_{eg} = \omega_{eg}/c$ match the Bloch components $k_x$ and $k_y$, respectively.
Consequently, different collective modes can be selectively excited by varying the orientation of the probe field, leading to directional coupling between the atomic array and the field.
This mechanism gives rise to a directional mode, whose property is determined by the incident angle $\theta$.
This mode has recently been utilized to map effective multichannel waveguide systems \cite{Multichannel}.

Here, we consider two incidence configurations for the probe light, as illustrated in Fig.~\figpanel{fig:1}{b}: incidence in the $x$-$z$ plane (corresponding to $k'_x=k'\sin{\theta}, k'_y=0$ and $k'_z=-k'\cos{\theta}$) and the $y$-$z$ plane (corresponding to $k'_x=0, k'_y=k'\sin{\theta}$ and $k'_z=-k'\cos{\theta}$).
For out-of-plane polarization, the atomic arrays exhibit four-fold rotational symmetry $C_4$, which results in identical collective responses for both incidence configurations.
Thus, we do not consider the orientation of the incident plane for out-of-plane polarization in the subsequent discussions.
In contrast, for the in-plane polarization, the response differs because the symmetry of the array reduces from $C_4$ to $C_2$ with respect to the polarization $\hat{\wp}_{x}$.
The band structures in Fig.~\figpanel{fig:2}{c} (corresponding to the $x$-$z$-plane incidence configuration) and Fig.~\figpanel{fig:2}{d} (corresponding to the $y$-$z$-plane incidence configuration) further reveal distinct collective responses of the atomic arrays to the probe fields with varying incident planes.

\begin{figure}[t]
\centering
\includegraphics[width=0.48\textwidth]{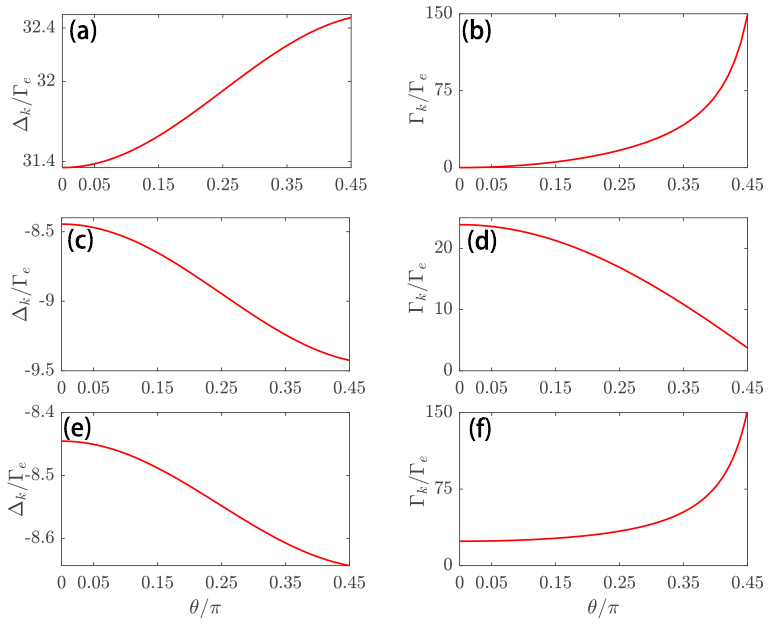}
\caption{Cooperative shift $\Delta_{k}$ and collective decay rate $\Gamma_{k}$ as functions of the incident angle $\theta$, for the out-of-plane polarization $\hat{\wp}= \hat{\wp}_{z}$ in (a) and (b), for the in-plane polarization $\hat{\wp}= \hat{\wp}_{x}$ with the probe field incident in the $x$-$z$ plane in (c) and (d), and for the in-plane polarization $\hat{\wp}= \hat{\wp}_{x}$ with the probe field incident in the $y$-$z$ plane in (e) and (f).
Other parameters are the same as those in Fig.~\ref{fig:2}.}
\label{Deltak_theta}
\end{figure}

For out-of-plane polarization, the directional collective mode excited by the probe field exhibits a pronounced blue-detuned shift with a cooperative shift of $\Delta_k \simeq 31.4\Gamma_e$, for probe incidence very close to normal.
Here, $\Delta_k \simeq 31.4\Gamma_e$ serves as the minimum reference value of the collective-mode dispersion (corresponding to the $\Gamma$ point), since an atomic array with out-of-plane polarization does not couple to a strictly normally incident field.
As the incident angle increases to $\theta = 0.45\pi$, $\Delta_k$ further increases by approximately $1.2\Gamma_e$, as shown in Fig.~\figpanel{Deltak_theta}{a}.
Meanwhile, as illustrated in Fig.~\figpanel{Deltak_theta}{b}, the collective decay rate remains close to zero near normal incidence and initially grows slowly, before rising sharply to $149\Gamma_e$ in a highly nonlinear manner as $\theta$ increases.
This behavior indicates a transition of the atomic array from a subradiant to a superradiant regime with increasing $\theta$, as the collective excitation mode gradually shifts away from the $\Gamma$ point and approaches the light cone, as shown in Fig.~\figpanel{fig:2}{a}.

For in-plane polarization, the collective mode exhibits distinct features of directional coupling, as shown in Fig.~\figpanel{Deltak_theta}{c--f}.
The optical response of the system shows a red-detuned shift, with an initial cooperative shift of $\Delta_k \simeq -8.4\Gamma_e$ when the probe field is incident perpendicular to the atomic array, as shown in Fig.~\figpanel{Deltak_theta}{c} for incidence in the $x$-$z$ plane and in Fig.~\figpanel{Deltak_theta}{e} for incidence in the $y$-$z$ plane.
As the incident angle increases to $\theta = 0.45\pi$, $\Delta_k$ exhibits a similar trend in both cases, decreasing to $-9.4\Gamma_e$ for the $x$-$z$-plane incidence configuration and to $-8.6\Gamma_e$ for the $y$-$z$-plane incidence configuration.
However, the collective decay rate $\Gamma_k$ displays distinct monotonic behaviors for the two incidence planes, as shown in Fig.~\figpanel{Deltak_theta}{d} and Fig.~\figpanel{Deltak_theta}{f}.
Starting from an initial value of $\Gamma_k\simeq 23.9\Gamma_e$ for $\theta=0$, the collective decay rate gradually decreases for the $x$-$z$-plane incidence configuration, indicating a change toward subradiant behavior.
Conversely, for the $y$-$z$ incidence case, $\Gamma_k$ increases markedly with $\theta$, signifying the emergence of pronounced superradiance due to strong coupling with free-space modes.

\begin{figure}[t]
\centering
\includegraphics[width=0.48\textwidth]{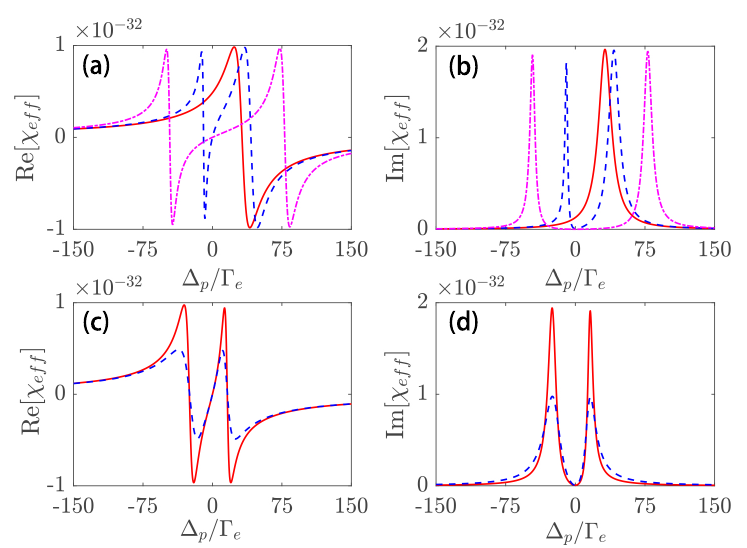}
\caption{(a) Real part $\mathrm{Re}\left[\chi_{\rm eff}\right]$ and (b) imaginary part $\mathrm{Im}\left[\chi_{\text{eff}}\right]$ of the effective susceptibility as functions of the probe detuning $\Delta_p$ for out-of-plane polarization $\hat{\wp} = \hat{\wp}_z$. The red solid, blue dashed, and magenta dot-dashed lines correspond to $\Omega_c = 0$, $20\Gamma_e$, and $60\Gamma_e$, respectively.
(c) $\mathrm{Re}\left[\chi_{\text{eff}}\right]$ and (d) $\mathrm{Im}\left[\chi_{\text{eff}}\right]$ as functions of $\Delta_p$ for in-plane polarization $\hat{\wp} = \hat{\wp}_x$, with $\Omega_c = 20\Gamma_e$.
The red solid and blue dashed lines denote probe incidence in the $x$-$z$ and $y$-$z$ planes, respectively.
Here $\theta = \pi/4$, $\Delta_c = 0$, and other parameters are the same as those in Fig.~\ref{fig:2}.}
\label{Fig.3}
\end{figure}

Figure~\ref{Fig.3} shows the real and imaginary parts of the probe susceptibility, illustrating the collective optical response of the atomic array with a three-level incidence configuration for both polarization cases.
When $\Omega_c = 0$, the three-level system reduces to a conventional two-level atomic array.
As shown in Fig.~\figpanel{Fig.3}{a} (referring to the dispersion) and Fig.~\figpanel{Fig.3}{b} (implying the absorption) for out-of-plane polarization, the spectra exhibit typical two-level characteristics; however, the resonance is blue-shifted from $\Delta_p = 0$ due to the cooperative shift (see Fig.~\figpanel{Deltak_theta}{a}).
With the application of a strong coupling field, the system enters the regime of EIT, where the single absorption peak splits into two, and an absorption-free valley (the EIT window) emerges between them near the corresponding $\Delta_k$ point.
Upon further increasing the coupling-field strength, the system exhibits Autler-Townes splitting.

For in-plane polarization, as shown in Fig.~\figpanel{Fig.3}{c} and \figpanel{Fig.3}{d}, the amplitude of the probe susceptibility for incidence in the $x$-$z$ plane is larger than that for incidence in the $y$-$z$ plane.
This behavior is consistent with the susceptibility given by Eq.~\ref{Eq:3} and reflects the pronounced difference in collective decay rates between the $x$-$z$-plane and $y$-$z$-plane incidence configurations (see Fig.~\figpanel{Deltak_theta}{d} and \figpanel{Deltak_theta}{f}).
Moreover, the positions and widths of the EIT windows and absorption peaks are well described by the decaying-dressed states (see Appendix~\ref{SMA}).

\section{Dual-reflection Spectra}\label{IV}

\begin{figure}[t]
\centering
\includegraphics[width=0.48\textwidth]{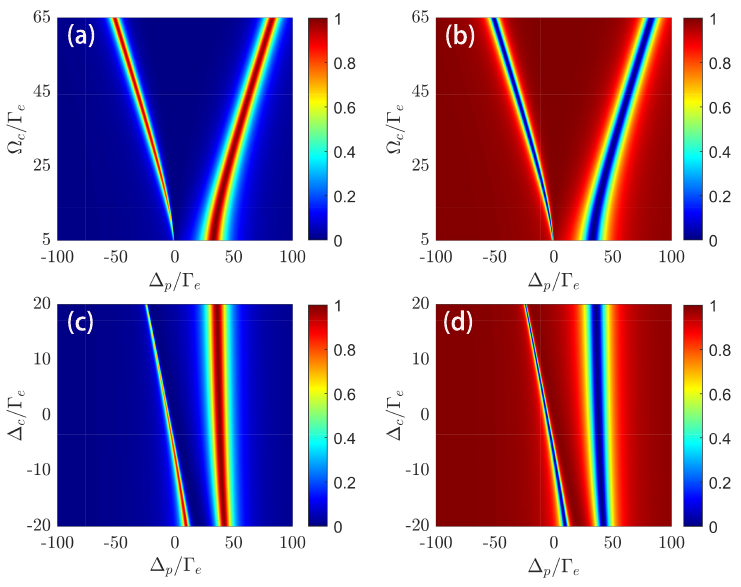}
\caption{(a) Reflectivity $R_{pp}$ and (b) transmissivity $T_{pp}$ as functions of the Rabi frequency $\Omega_c$ and probe detuning $\Delta_p$ with $\Delta_c = 0$.
(c) Reflectivity $R_{pp}$ and (d) transmissivity $T_{pp}$ as functions of coupling detuning $\Delta_c$ and probe detuning $\Delta_p$ with $\Omega_c = 15\Gamma_e$.
Here $\hat{\wp} = \hat{\wp}_z$, $\theta = \pi/4$, and other parameters are the same as those in Fig.~\ref{fig:2}.}
\label{fig.3}
\end{figure}

Based on the tunable collective response discussed above, we further investigate the manipulation of reflection and transmission via system parameters.
Here, we consider a $p$-polarized probe field incident on the array.
As shown in Fig.~\figpanel{fig.3}{a} and \figpanel{fig.3}{b}, both the reflection and transmission spectra vary with the intensity of the coupling field.
The transparency window gradually broadens as $\Omega_c$ increases.
It is worth noting that the reflection spectra exhibit two distinct reflection bands: a narrow band in the red-detuned region and a broader band in the blue-detuned region.
This property enables the design of a collective atomic mirror with a dual-reflection-band response in the superradiance region of the collective mode, with the operational bandwidths tunable through the adjustment of the coupling-field amplitude.
Specifically, the narrow band broadens while the wide band becomes slightly narrower as $\Omega_c$ increases.
As shown in Fig.~\figpanel{fig.3}{c} and Fig.~\figpanel{fig.3}{d}, when the coupling detuning $\Delta_c$ increases from the red- to the blue-detuned side, the bandwidth of the left (narrow) reflection peak decreases significantly, suggesting the possibility of realizing an ultranarrow-band atomic mirror, while the right (broader) peak exhibits slight broadening.

Such a dual-band atomic mirror may also exhibit selectivity with respect to the polarization state of the probe light.
For out-of-plane polarization, in both incidence configurations (the $x$-$z$ and $y$-$z$ planes), the electric field of the $s$-polarized probe remains orthogonal to the atomic polarization.
This makes the atomic array fully transmissive to the $s$-polarized light ($T_{ss}\equiv 1 $ and $R_{ss}\equiv 0 $).
In contrast, the $p$-polarized probe can couple to the atomic dipoles and then excite the collective response of the array.
This polarization-dependent behavior enables the implementation of a polarization beam splitter~\cite{beamsplitter,cai2015ultra}.
For in-plane polarization, when the probe field is incident in the $x$-$z$ plane ($y$-$z$ plane), only the $p$-polarized ($s$-polarized) component can induce the collective response, allowing this atomic-mirror scheme to function as a polarization filter~\cite{beamsplitter,guo2016broadband} for the two orthogonal incidence planes or as an optical diode for linearly polarized light~\cite{diode}.

\begin{figure}[t]
\centering
\includegraphics[width=0.48\textwidth]{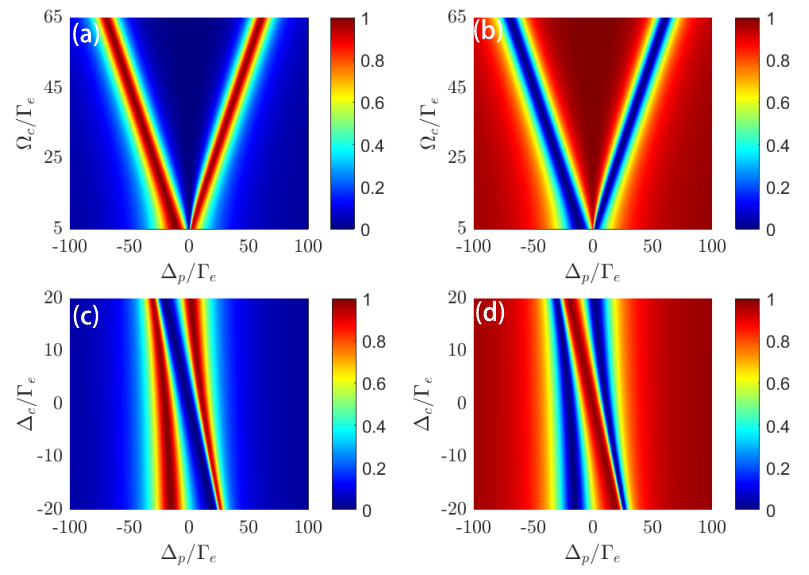}
\caption{(a) Reflectivity $R_{ss}$ and (b) transmissivity $T_{ss}$ as functions of the Rabi frequency $\Omega_c$ and probe detuning $\Delta_p$ with $\Delta_c = 0$.
(c) Reflectivity $R_{ss}$ and (d) transmissivity $T_{ss}$ as functions of the coupling detuning $\Delta_c$ and probe detuning $\Delta_p$ with $\Omega_c = 15\Gamma_e$.
The probe field is incident in the $y$-$z$ plane.
Here $\hat{\wp} = \hat{\wp}_x$, $\theta = \pi/4$, and other parameters are the same as those in Fig.~\ref{fig:2}.
}
\label{fig6}
\end{figure}

As shown in Fig.~\figpanel{fig6}{a} and \figpanel{fig6}{b}, where an $s$-polarized probe field is incident on the atomic array with in-plane polarization $\hat{\wp} = \hat{\wp}_x$, a substantially broader reflection band is obtained compared with the results in Fig.~\ref{fig.3}, accompanied by a reversal of the asymmetric features of the two reflection peaks.
This behavior is advantageous for polarization-filter applications.
The enhanced bandwidth arises from the parallelism between the probe field and atomic polarizations, which leads to stronger coupling.
Moreover, the two reflection peaks exhibit clearly opposite trends in their bandwidths as the coupling-field frequency varies, enabling on-demand switching between spectral profiles of the filter.
In contrast to the case of incidence in the $x$-$z$ plane, the $p$-polarized probe displays similar spectral features, with only a slight narrowing of the bandwidth.

\begin{figure}[t]
\centering
\includegraphics[width=0.48\textwidth]{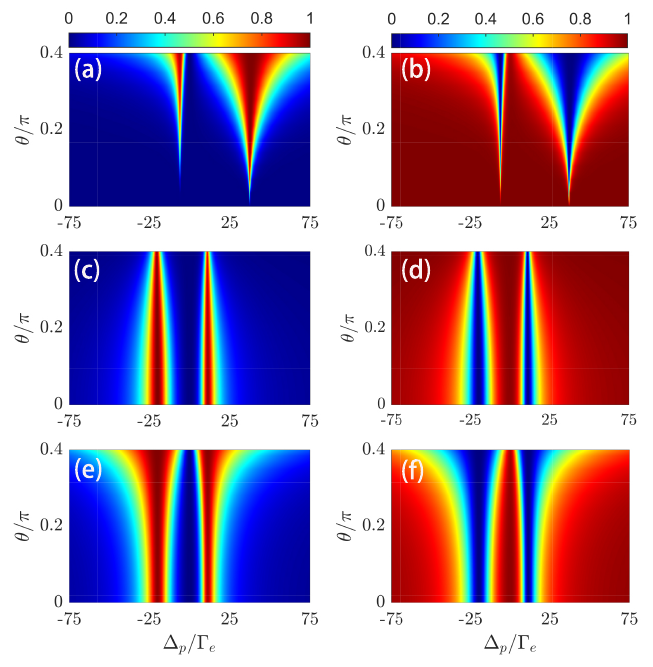}
\caption{(a) Reflectivity $R_{pp}$ and (b) transmissivity $T_{pp}$ as functions of the incident angle $\theta$ and probe detuning $\Delta_p$ for out-of-plane polarization $\hat{\wp} = \hat{\wp}_{z}$.
For in-plane polarization $\hat{\wp} = \hat{\wp}_{x}$, (c) reflectivity $R_{pp}$ and (d) transmissivity $T_{pp}$ are shown for incidence in the $x$-$z$ plane, and (e) reflectivity $R_{ss}$ and (f) transmissivity $T_{ss}$ are shown for incidence in the $y$-$z$ plane.
Here $\Omega_c = 15\Gamma_e$, and $\Delta_c = 0$, and other parameters are the same as those in Fig.~\ref{fig:2}.
}
\label{fig7}
\end{figure}

Figure~\ref{fig7} shows that the dual-band atomic mirror can be realized not only under normal incidence ($\theta = 0$) but also for oblique incidence.
For out-of-plane polarization, as the incident angle $\theta$ increases, the bandwidths of both reflection peaks broaden significantly, and the peak reflectivity increases, as shown in Figs.~\ref{fig7}{a} and \ref{fig7}{b}.
For in-plane polarization with incidence in the $x$-$z$ plane, as shown in Figs.~\ref{fig7}{c} and \ref{fig7}{d}, the bandwidth for the $p$-polarized light decreases with increasing $\theta$, whereas for incidence in the $y$-$z$ plane, the bandwidth of the $s$-polarized light exhibits the opposite trend, as shown in Figs.~\ref{fig7}{e} and \ref{fig7}{f}.
Overall, Fig.~\ref{fig7} indicates that the atomic mirror can operate efficiently over a wide angular range, maintaining high reflectivity ($R > 0.97$).

\begin{figure}[t]
\centering
\includegraphics[width=0.48\textwidth]{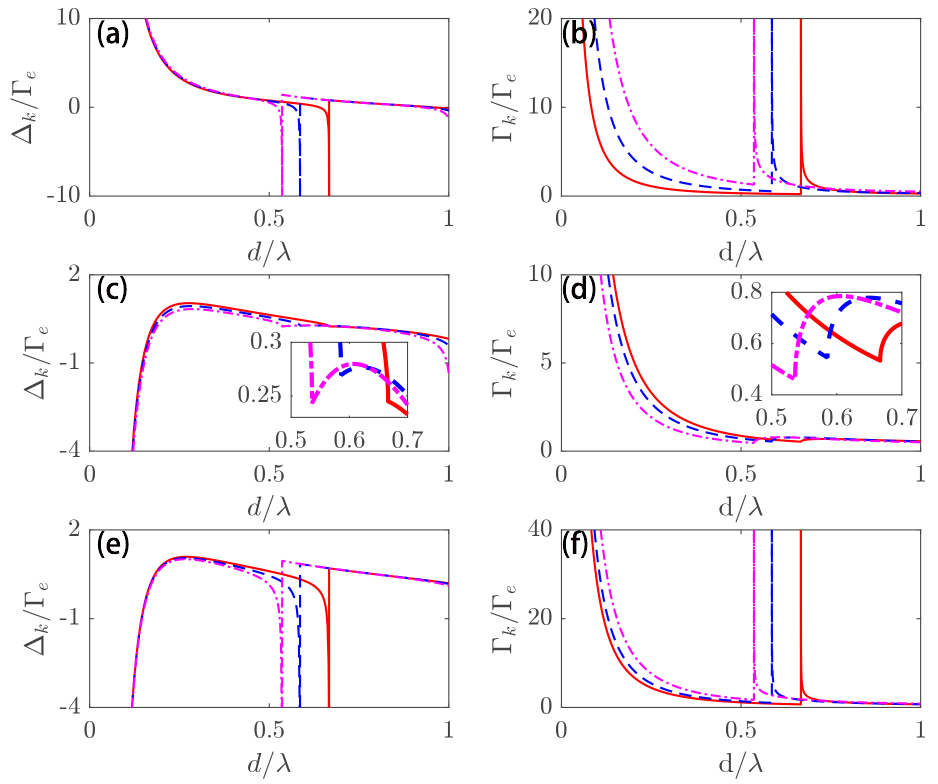}
\caption{Cooperative shift $\Delta_{k}$ and collective decay rate $\Gamma_{k}$ as functions of the lattice constant $d$ for out-of-plane polarization $\hat{\wp}=\hat{\wp}_{z}$ in (a) and (b), and for in-plane polarization $\hat{\wp}=\hat{\wp}_{x}$ with incidence in the $x$-$z$ plane in (c) and (d), and in the $y$-$z$ plane in (e) and (f).
The red solid, blue dashed, and magenta dot-dashed lines correspond to incident angles $\theta=\pi/6,\ \pi/4,\ \pi/3$, respectively.
The inset shows an enlarged view.
Other parameters are the same as those in Fig.~\ref{fig:2}.}
\label{fig8}
\end{figure}

In general, the dual-band atomic-mirror scheme can operate over a relatively wide parameter range as long as the system remains in the superradiant regime.
However, the other-order diffraction may weaken the zero-order diffraction (the normal reflection)~\cite{PRL2017scattering}.
We therefore provide the general condition for the appearance of diffraction over a broad range of incident angles.
When the lattice constant $d$ exceeds $\lambda/2$, Bragg scattering opens additional dissipation channels (diffraction order), leading to emission into directions other than the primary reflection and transmission.
This behavior can be verified by analyzing the collective modes.

The components of the discrete Fourier transform of the dyadic Green's function are given by
\begin{align}
    G_{\mathbf{k'}_{\parallel}}^{ij}=\frac{i}{2d^2}\sum_m \frac{\delta_{ij}-(\mathbf{q}_m-\mathbf{k'}_{\parallel})_i(\mathbf{q}_m-\mathbf{k'}_{\parallel})_j/k'^2}{\sqrt{k'^2-|\mathbf{q}_m-\mathbf{k'}_{\parallel}|^2}} \label{G_k},
\end{align}
with $i,j=x,y$, where $\mathbf{q}_m$ are the reciprocal lattice vectors satisfying $\mathbf{q}_m \cdot \mathbf{r}_n = 2\pi n$ with integer $n$.
For a square lattice, $\mathbf{q}_m = (2\pi/d)(m_x\mathbf{e}_x + m_y\mathbf{e}_y)$.
A key condition for superradiant modes localized within the light cone is that the square root in the denominator remains real.
Physically, this corresponds to requiring a real $k'_z$.
A complex $k'_z$ describes an evanescent, spatially localized wave, which lies outside the light cone.

For the $y$-$z$-plane incidence configuration, $k'_x=0$ and $k'_y = k' \sin\theta$, we have
\begin{align}
k'^2 > \left(\frac{2\pi}{d}m_x\right)^2 + \left(\frac{2\pi}{d}m_y - k' \sin\theta\right)^2 .
\end{align}
For $d < \lambda/2$, only the zeroth-order diffraction ($m_x = m_y = 0$) contributes to radiation and damping.
For $\lambda > d > \lambda/2$, an additional diffraction order $(m_x,m_y) = (0,1)$ emerges when
\begin{align}
d = \lambda/(1+\sin\theta).
\end{align}
For the $x$-$z$-plane incidence configuration ($k'_x = k'\sin\theta$ and $k'_y = 0$), the additional diffraction order becomes $(m_x,m_y) = (1,0)$.
These higher-order diffractions manifest as distinct peaks in the collective mode spectrum~\cite{PRL2017scattering}, as shown in Fig.~\ref{fig8}, with peak positions located at $d = \lambda/(1+\sin\theta)$.

Notably, for in-plane polarization and incidence in the $x$-$z$ plane, when the probe field is aligned parallel to the atomic polarization, the peak associated with the additional diffraction order is suppressed and appears instead as an inflection point in the collective response, as shown in Fig.~\figpanel{fig8}{c} and ~\figpanel{fig8}{d}.
This behavior follows from Eq.~(\ref{G_k}).
For $k'_x=k'\sin\theta$, $k'_y=0$,
\begin{align}
    G_{\mathbf{k'}_{\parallel}}^{xx}&=\frac{i}{2d^2}\sum_m \frac{(1-(2\pi m_x/d-k'_x)^2/k'^2)}{\sqrt{k'^2-|\mathbf{q}_m-\mathbf{k'}_{\parallel}|^2}} \nonumber\\
    &=\frac{i}{2d^2}\sum_m \frac{(1-(\lambda m_x/d-\sin\theta)^2)}{\sqrt{k'^2-|\mathbf{q}_m-\mathbf{k'}_{\parallel}|^2}}.
\end{align}
At $d = \lambda/(1+\sin\theta)$, the contribution from the diffraction order $(m_x,m_y) = (1,0)$ vanishes, i.e., $G_{\mathbf{k}}^{xx}(m_x,m_y)=0$.
Consequently, propagation along that direction is suppressed, producing a retro-reflection effect in which most of the incident light is redirected into the zeroth-order diffraction channel~\cite{PRL2017scattering}.

Our proposal is based on cold atomic systems, where ultracold atoms are trapped in subwavelength periodic potentials formed by optical lattices~\cite{schafer2020tools,kumar2018sorting} or optical tweezers~\cite{glicenstein2020collective,sheng2022defect}.
In these systems, the interactions depend only on the relative positions and polarization orientations of the atoms (see Eq.~(\ref{H_eff}).
The polarization vectors are defined with respect to a quantization axis about which the optically excited orbitals are circularly polarized; experimentally, this axis is set by an external electric or magnetic field.
Our approach can also be implemented in other platforms, including plasmonic nanoparticles~\cite{SMHPo1}, silicon carbide nanoparticles~\cite{NPs}, transition metal dichalcogenides~\cite{TMD1, TMD2}, and quarter-wavelength resonators~\cite{QWR}.

\section{Conclusion} \label{V}
We have investigated the cooperative optical response of a two-dimensional subwavelength atomic array driven in a three-level ladder configuration. The interplay between collective radiative modes and EIT gives rise to a dual-band atomic mirror featuring two reflection resonances with distinct linewidths. These bands originate from cooperative frequency shifts and collective decay rates, and their spectral positions and bandwidths can be tuned through the incident angle, dipole orientation, lattice constant, and coupling-field parameters. Both out-of-plane and in-plane polarizations support strong directional coupling but exhibit different spectral characteristics, enabling the array to operate as a polarization beam splitter or filter. Additional tunability is provided by the EIT and Autler-Townes regimes, allowing the dual-band response to be reconfigured in situ through adjustments to the coupling-field strength or the incidence geometry.
We further analyzed the conditions under which higher-order diffraction appears, thereby identifying the parameter regime in which the high-efficiency atomic mirror retains full tunability.

\begin{figure}[t]
\centering
\includegraphics[width=0.48\textwidth]{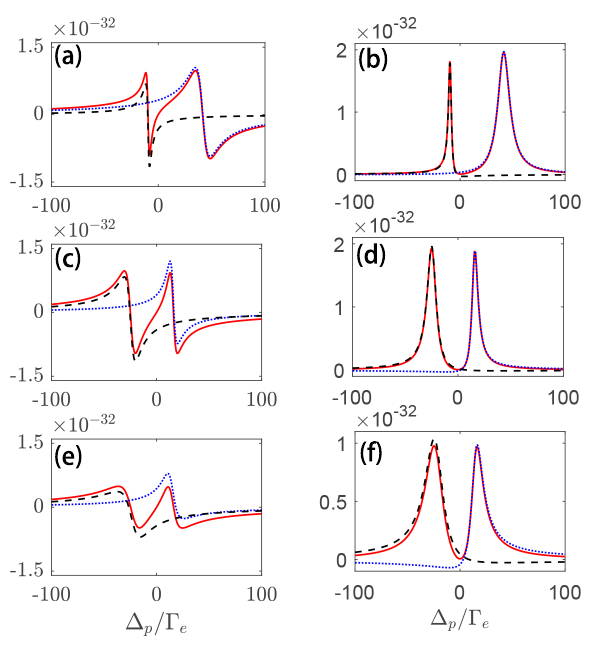}
\caption{Real parts of $\chi_{\rm eff}$, $\beta_1$, and $\beta_2$ as functions of the probe detuning $\Delta_p$ for out-of-plane polarization $\hat{\wp} = \hat{\wp}_z$ in (a); for in-plane polarization $\hat{\wp} = \hat{\wp}_x$ with probe incidence in the $x$-$z$ plane in (c); and for in-plane polarization $\hat{\wp} = \hat{\wp}_x$ with probe incidence in the $y$-$z$ plane in (e).
Imaginary parts of $\chi_{\rm eff}$, $\beta_1$, and $\beta_2$ as functions of $\Delta_p$ are shown for the same polarization and incidence configurations in (b), (d), and (f), respectively.
The red solid, blue dashed, and black dotted lines correspond to $\chi_{\rm eff}$, $\beta_1$, and $\beta_2$, respectively.
Here $\theta = \pi/4$, $\Omega_c = 20\Gamma_e$, $\Delta_c = 0$, and other parameters are the same as those in Fig.~\ref{fig:2}.}
\label{decaying-dressed}
\end{figure}

Our scheme provides a more accessible and versatile method for manipulating cooperative optical behavior than schemes based on strong Rydberg interactions.
Because it relies only on single-excitation collective physics combined with EIT control, it avoids the stringent requirements associated with Rydberg blockade and long-range interactions.
The tunable cooperative response, together with the dual-band reflection capability, provides a practical foundation for integrating atomic-mirror elements into all-optical networks and developing reconfigurable photonic components.
These results highlight the broader potential of ordered atomic arrays as a flexible and experimentally viable platform for tunable photonic functionalities and more advanced applications in quantum optics and integrated nanophotonics.

\vspace{1em}

\section*{ACKNOWLEDGMENTS}
This work is supported by the Science Foundation of the Education Department of Jilin Province (No. JJKH20250301KJ) and Jilin Scientific and Technological Development Program (No. 20240101328JC).

\appendix
\section{Responses via decaying-dressed states}\label{SMA}

The effective susceptibility $\chi_{\rm eff}$ in Eq.~\ref{Eq:3} can be expressed as
\begin{align}
\chi_{\rm eff}&=\beta_1(\Delta_p)+\beta_2(\Delta_p),\nonumber\\
  \beta_1(\Delta_p)&=\frac{\left | \boldsymbol{\mu}_{eg} \right| ^2 }{\hbar}\frac{\xi-\Delta_+}{\Delta_+-\Delta_-} \cdot\frac{1}{\Delta_p-\Delta_+},\nonumber\\
  \beta_2(\Delta_p)&=\frac{\left | \boldsymbol{\mu}_{eg} \right| ^2 }{\hbar}\frac{\Delta_--\xi}{\Delta_+-\Delta_-} \cdot\frac{1}{\Delta_p-\Delta_-},
  \label{Eq:7}
  \end{align}
presenting two response functions that illustrate the transition from the corresponding decaying-dressed states to the ground states.
Then, we plot the effective susceptibility $\mathrm{Im}(\beta_1)$ together with the response functions $\beta_1$ and $\beta_2$ (corresponding to the decaying-dressed states) for out-of-plane polarization and for in-plane polarization with incidence in the $x$-$z$ and $y$-$z$ planes, as shown in Fig.~\ref{decaying-dressed}.
In this representation, the absorption features in $\mathrm{Im}(\beta_1)$ and $\mathrm{Im}(\beta_2)$ exhibit Lorentzian line shapes.


%
\end{document}